\documentclass[runningheads]{llncs}

\usepackage[T1]{fontenc}
\usepackage[utf8]{inputenc}
\usepackage{amsmath}
\usepackage{amssymb}
\usepackage{graphicx,verbatim}
\usepackage{booktabs}
\usepackage[table]{xcolor}
\usepackage[
    colorlinks=true,
    citecolor=green,
    linkcolor=black,
    urlcolor=blue
]{hyperref}
\usepackage{orcidlink}
\begin{document}

\title{Do Medical Vision Models Reason About\\ Anatomy? Probing the Spatial Inductive Biases\\ of Learned Visual Representations}
\titlerunning{Do Medical Vision Models Reason About Anatomy?}

\author{Naren Akash\inst{1, 2}\orcidlink{0000-0002-5218-8434} \and
Neeraja Ramanan\inst{1, 2}}
\authorrunning{N. Akash et al.}
\institute{Center for Visual Information Technology, IIIT Hyderabad, Hyderabad, India \and
Center for AI and Innovation, AIG Hospitals, Hyderabad, India
\email{naren.akash@research.iiit.ac.in}\\
}


\maketitle


\begin{abstract}

Interpreting a CT scan means comparing structures on either side, judging how far apart
organs sit, and knowing where each one belongs. Medical vision encoders are evaluated on
diagnostic accuracy, or through assembled multimodal systems where a failure is hard to
attribute, so it remains unclear whether their representations support any of this. We
construct SPAR-Bench, eight probes over multi-organ abdominal CT that separate coordinate
localization, relational reasoning, and spatial queries, and apply them to five
architectural configurations and three medical foundation models, frozen and finetuned.
Probes that ask for a comparison within the slice stay at chance, and neither pretraining scale, finetuning, nor architecture closes the gap. Probes
that appear solved in domain fall to chance under zero-shot transfer, indicating that
their accuracy reflects recall of canonical anatomy rather than computation over the
image. Reading the same frozen features with a pooled head rather than the full set of tokens
moves relational recovery from 0.7\% to 67.8\%, so pooled probing understates what a
representation holds. Questions the encoders answer well are answered
at chance by four open-weight MLLMs. Our results suggest these encoders carry a map of
where organs usually lie, and little of the machinery for comparing structures within a
particular patient. Code and data will be available at
\href{https://spar-bench.github.io}{\texttt{spar-bench.github.io}}.


\end{abstract}
\section{Introduction}

Radiological diagnosis depends on the spatial coordination of anatomical structures.
A finding's clinical significance is often determined by its position within a
\textit{canonical spatial map}, its proximity to adjacent landmarks, and its
appearance relative to the contralateral side. Implicit in expert interpretation is
the utilization of this map: staging a pancreatic mass requires quantifying its
contact with the superior mesenteric artery~\cite{alhawary2014pancreatic}.

Models are scored on the diagnosis they output. A diagnostic label, however, sits at
the end of a chain of simpler judgements, and reaching it requires composing several
of them, including the presence of some features and the confirmed absence of others.
Screening for renal artery stenosis turns on whether the two kidneys differ in
size~\cite{artyszuk2022threshold}, which requires locating both, measuring each, and
comparing the two. Accuracy on the final label scores the composition and leaves the
elements it was built from untested, so a model can reach the correct finding through
texture or context while representing none of the underlying
geometry~\cite{geirhos2020shortcut}. The same gap appears in evaluations of medical
multimodal LLMs, which report failures on image orientation and on judging which of
two slices lies closer to a landmark~\cite{bigverdi2025medblink}, and answers that flip
between visually similar image pairs~\cite{sepehri2025mediconfusion}. Both evaluate
assembled systems, in which a wrong answer can originate in the encoder, the
projection layer, or the language model. We instead probe the elements, and probe
them in the vision encoder alone.

Prior work has characterized the spatial properties of vision architectures primarily
from two directions. First, representation analyses show that Vision Transformers
(ViTs) preserve spatial information in tokens across layers more uniformly than
CNNs~\cite{raghu2021vit}, and show superior robustness to patch-level
permutations~\cite{naseer2021intriguing}. Second, self-supervised methods in medical
imaging largely utilize spatial disruption, such as shuffling, masking, or rearranging
patches, as pretext signals to encourage anatomical representation learning. Both
lines focus on representation preservation or reconstructive robustness rather than
functional reasoning. Existing benchmarks evaluate basic perception in medical
multimodal LLMs~\cite{bigverdi2025medblink,sepehri2025mediconfusion} or analyze natural
vision models~\cite{chen2024spatialvlm,ma20243dsrbench}, but none quantify whether a
vision encoder can compute over anatomical spatial structure.

We introduce \textbf{SPAR-Bench} (Spatial Anatomical Reasoning Benchmark), which probes
spatial competency in vision encoders through three hierarchical levels. Localization tests whether models can predict organ centroid coordinates.
Relational Reasoning tests whether models can reconstruct shuffled anatomical tiles.
Spatial Queries test whether models can answer six question types spanning relative
position, symmetry, size, distance, and counting. Using multi-organ abdominal CT as a
testbed, we evaluate five architectural configurations
and three medical foundation models as frozen encoders with lightweight trainable heads
and in fine-tuned settings. We further test zero-shot
transfer to TotalSegmentator and pose the same spatial questions to four open-weight
MLLMs.

We make four contributions. (i) We introduce SPAR-Bench, eight probes derived
automatically from segmentation masks that separate localization, relational reasoning,
and spatial queries. (ii) We show that bilateral symmetry stays at chance across all 13
configurations, under zero-shot transfer, and for the MLLMs we test, while relative
position reaches 99\% on WORD and falls to chance out of domain, which suggests the
in-domain score reflects recall of canonical anatomy rather than a computation over the
image. (iii) We show that reading the same frozen features with a pooled head rather than the
full token sequence moves relational recovery from 0.7\% to 67.8\%, which affects how
frozen-model evaluations should
be read. (iv) We evaluate encoders and assembled MLLMs on the same questions and find
that questions the encoders answer well are answered at chance by the MLLMs.

\section{Related Works}

\noindent \textbf{Spatial Reasoning over Anatomy.} Self-supervised methods have used
spatial structure as a training signal: jigsaw solving~\cite{noroozi2016jigsaw}, Rubik's
cube recovery~\cite{zhuang2019rubiks}, and Models Genesis~\cite{zhou2019models} learn
through spatial disruption and restoration, while Bai et
al.~\cite{bai2019selfsupervised} use position prediction to supervise cardiac MRI
segmentation. Benchmarks exist for natural
vision~\cite{chen2024spatialvlm,ma20243dsrbench} and medical
MLLMs~\cite{bigverdi2025medblink,sepehri2025mediconfusion}, but none tests whether vision
encoders reason about anatomical space: localizing organs, reconstructing arrangements,
or comparing bilateral pairs.\\

\noindent \textbf{Model Capacity vs.\ Representation.} Medical foundation models such as
RadDINO~\cite{perezgarcia2025raddino} and BiomedCLIP~\cite{zhang2023biomedclip} are
evaluated through classification transfer, which can succeed on local texture without
spatial reasoning. UniBench~\cite{altahan2024unibench} found scaling provides no benefit
for spatial understanding across 60 natural-vision models, raising whether medical
pretraining at scale resolves this. CNNs and ViTs handle spatial information
differently~\cite{raghu2021vit,naseer2021intriguing}, but whether this affects functional
spatial reasoning has not been tested. Encoder--decoder architectures such as
TransUNet~\cite{chen2024transunet}, Swin-UNet~\cite{cao2022swinunet},
UNETR~\cite{hatamizadeh2022unetr}, and nnU-Net~\cite{isensee2021nnunet} recover spatial
detail through skip-decoded fusion, which mixes encoder content with the decoder's
contribution and prevents attributing a capability to either part. We therefore scope
SPAR to encoder representations read by classification-style heads, matching how these
encoders are consumed in multimodal systems.
\section{Spatial Anatomical Reasoning (SPAR) Benchmark}

\subsection{Benchmark Construction}

We construct SPAR-Bench from the Whole Abdominal Organ Dataset (WORD) ~\cite{luo2022word}: 150 abdominal CT volumes with voxel-level masks for 16 organs, split into 100 train, 20 validation, and 30 test. WORD includes 20 volumes from the Liver Tumor Segmentation (LiTS) benchmark~\cite{bilic2023lits} with tumor annotations, totaling 170 volumes.
We extract 2D axial slices from the inter-quartile range of each organ's axial extent to avoid partial visibility. For Levels 1 and 2, we retain only slices with at least five organs present. Each organ is cropped to its tightest bounding box and resized to 224$\times$224; missing organs are black tiles. For Level 3, whole axial slices are resized to 224$\times$224 with CT windowing (level 50, width 400). Cropping at Levels 1 and 2 removes surrounding context, so a model that places a tile correctly must have encoded position from the organ's own appearance rather than from neighbouring structures.

\noindent \textbf{What we mean by spatial reasoning.} We use the term for tasks that compose two or more spatial measurements within a single image and cannot be answered from a fixed anatomical prior. Bilateral symmetry meets this criterion, since kidney area ratios vary across patients. Permutation recovery meets it, since each tile's slot depends on the other tiles present. Relative position and distance do not meet it as strictly, and we return to this below. The eight probes span absolute position, pairwise relations, and measurement within a single image, and all are derivable from segmentation masks without manual annotation. Probes needing 3D context, such as containment or vessel adjacency, are outside what 2D axial slices support.\\

\noindent \textbf{Level 1: Localization (Coordinate Regression).} The input is 16 organ tiles per slice. The model predicts normalized (x, y) coordinates for each organ slot. Targets are organ centroids in the original slice frame, normalized to slice dimensions. Training uses MSE loss masked by organ presence. We report Mean Euclidean Distance (MED) in pixels and Success Detection Rate (SDR) at a 10-pixel threshold.\\

\begin{figure}[!t]
\centering
\includegraphics[width=\textwidth]{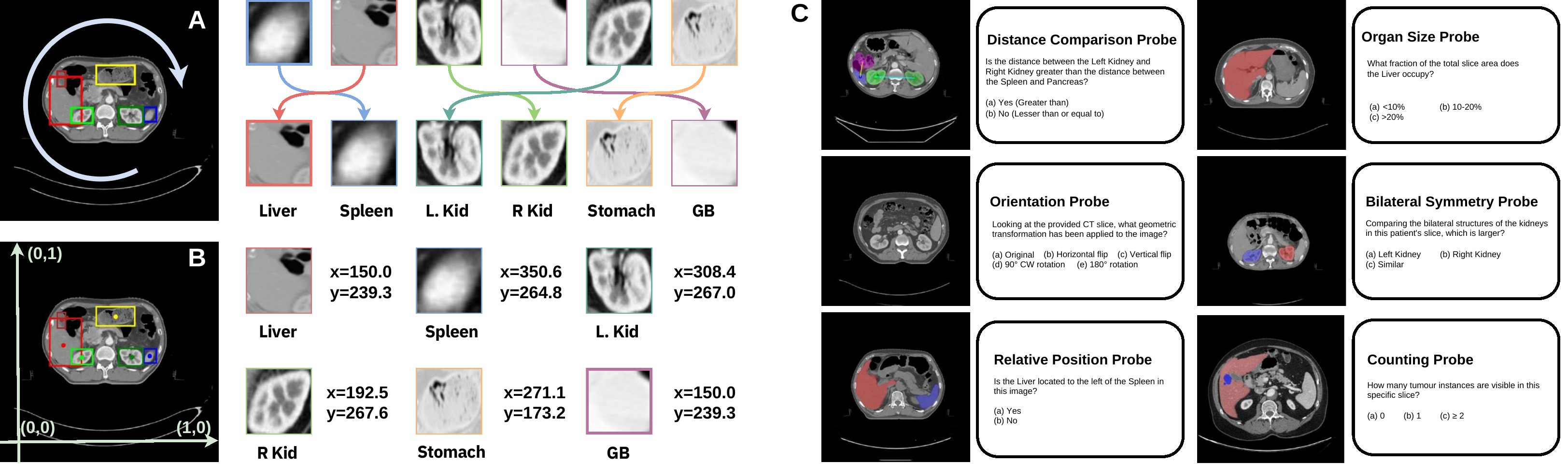}
\caption{SPAR-Bench. \textbf{(A)} Organs are cropped into 16 tiles, shuffled for Level~2. \textbf{(B)} Level~1 maps tiles to normalized coordinates. \textbf{(C)} Level~3 pairs whole slices with six probe types.}
\label{fig:teaser}
\end{figure}

\noindent \textbf{Level 2: Relational Reasoning (Permutation Prediction).} The same 16 organ tiles are randomly shuffled. The original ordering places each organ in the grid cell matching its centroid, so the model must recover a spatial arrangement, not a fixed class order. The model outputs a 16×16 doubly-stochastic permutation matrix via Sinkhorn normalization~\cite{sinkhorn1967doubly}, with the Hungarian algorithm~\cite{kuhn1955hungarian} producing a hard assignment at inference. Missing organ slots are ignored, so permutation is evaluated only over present organs. We report patch accuracy over tiles and permutation accuracy over fully correct sequences.\\

\noindent \textbf{Level 3: Spatial Queries (Six Probes).} The input is a whole 2D CT slice (224$\times$224, windowed), paired with a question embedding that routes the sample to one of six heads: \textbf{(i) Relative Position} (2 classes): Is organ A left or right of organ B? \textbf{(ii) Orientation} (5 classes): Is the slice in original orientation, horizontally flipped, vertically flipped, rotated 90°, or rotated 180°? \textbf{(iii) Bilateral Symmetry} (3 classes): Are the kidneys symmetric, left-larger, or right-larger? Labels come from per-slice cross-sectional area ratios on slices with both kidneys present, with the symmetric band set to 0.95--1.05. \textbf{(iv) Organ Size} (3 classes): What fraction of the slice does the organ occupy: $<$10\%, 10--20\%, or $>$20\%? Tested on liver. \textbf{(v) Distance} (2 classes): Is the distance between one organ pair greater or less than another? \textbf{(vi) Counting} (3 classes): How many tumor instances are present: 0, 1, or $\geq$2? Uses the 20 LiTS volumes with tumor annotations. Training uses cross-entropy loss per head. We report per-probe accuracy and macro-average across all six probes. Three probes carry controls against memorized answers. Relative position presents each slice in original and horizontally flipped form, so the label flips with the image and recall of the canonical arrangement scores at chance. Distance compares organ pairs that change across slices, so no single ordering applies. Symmetry balances its three classes by undersampling, so a constant answer cannot score above chance. These controls constrain what a model can exploit within WORD but not regularities shared across abdominal CT, which Section~\ref{sec:external} addresses.

\subsection{Models and Probing}

We factorize backbone (CNN vs.\ ViT) and reasoning head (MLP vs.\ Transformer) in a 2$\times$2 design, plus a fifth unified configuration, using ResNet-50~\cite{he2016resnet} and ViT-B/32~\cite{dosovitskiy2021vit}, both ImageNet-initialized. Hybrid architectures entangle inductive bias with pretraining scale, augmentation, and decoder design, so comparing them would confound the factor of interest. ResNet-50 and ViT-B/32 differ principally in how spatial structure is imposed, through convolutional locality or global attention over patch tokens, following established practice in representation analysis~\cite{raghu2021vit,naseer2021intriguing}. MLP against Transformer heads is likewise the minimal contrast between pooling a representation and computing over it. These are not competitive baselines; they establish what is achievable under modest supervision, against which we read foundation model performance.
MLP heads read pooled features, the CLS token for ViT and average-pooled features for CNN. Transformer heads read the full token sequence. ViT-Combined patchifies all 16 tiles jointly with tile-specific and positional embeddings. Question conditioning is concatenated for MLP heads and appended as a token for Transformer heads. We further replace the backbone with three medical foundation models: RadDINO~\cite{perezgarcia2025raddino} (radiograph self-supervised), BiomedCLIP~\cite{zhang2023biomedclip} (15M image-text pairs), and SAM-Med2D~\cite{cheng2023sammed2d} (segmentation-pretrained), attaching the same task heads. Frozen models train only the task heads. For RadDINO we also evaluate full fine-tuning, to test whether unfreezing recovers what freezing does not.

All models are trained with Adam, learning rate $10^{-4}$ (logarithmic search over $10^{-3}$ to $10^{-6}$), batch size 16, maximum 50 epochs with early stopping on validation loss (patience 20). Level 2 uses Sinkhorn $\tau$ = 0.3 with 20 iterations and smoothed permutation loss ($\varepsilon$ = 0.1). Each experiment is repeated across three seeds (42, 123, 1234); we report mean and standard deviation. Training uses a single NVIDIA RTX 4090, and all configurations train on identical splits, independently per level.

\subsection{External Evaluation and SPAR-VQA}
\label{sec:external}

A regularity common to abdominal CT would survive our probe-level controls, so we evaluate every trained configuration on TotalSegmentator~\cite{wasserthal2023totalsegmentator} without retraining, deriving labels by the procedure of Section~3.1. A probe answered by recall of canonical anatomy should degrade on a cohort with different acquisition and patients.
We also pose the Level 3 questions to four open-weight MLLMs: LLaVA-1.5-7B~\cite{liu2024llava}, Qwen2-VL-2B-Instruct and Qwen2-VL-7B-Instruct~\cite{wang2024qwen2vl}, and MedGemma-4B-IT~\cite{sellergren2025medgemma}. Each probe is phrased through multiple natural-language templates that differ in wording but ask the same question, sampled uniformly per item with the seed controlling the sampling. We evaluate on a resampled subset of the WORD test slices, giving 1{,}162 relative position, 1{,}050 orientation, 444 symmetry, 712 organ size, and 170 distance items; tumor counting is omitted. We do not vary answer-option order, so those effects remain untested.
\section{Results and Discussion}
\begin{table}[!t]
\centering
\caption{Spatial reasoning performance, mean over three seeds (two for TotalSegmentator
localization). MED is Mean Euclidean Distance in px; Perm is full permutation accuracy;
RelPos, Orient and Symm are the relative position, orientation and bilateral symmetry
probes. Accuracies in \%. $^\dagger$Finetuned end-to-end. \textbf{Bold} marks the best value per
column on WORD; no value is marked for symmetry, where all configurations sit at chance.}
\label{tab:main_results}
\scriptsize
\setlength{\tabcolsep}{3pt}
\begin{tabular}{lccccc|ccccc}
\toprule
& \multicolumn{5}{c|}{WORD (in domain)} & \multicolumn{5}{c}{TotalSegmentator (zero-shot)} \\
\cmidrule(lr){2-6} \cmidrule(lr){7-11}
Model & MED$\downarrow$ & Perm & RelPos & Orient & Symm & MED$\downarrow$ & Perm & RelPos & Orient & Symm \\
\midrule
\multicolumn{11}{l}{\textit{Trained from scratch}} \\
CNN-MLP          & \textbf{8.95} & 64.7 & 99.8 & 93.6 & 29.9 & 70.9 & 16.9 & 68.8 & \phantom{0}9.1 & 30.9 \\
CNN-Trans        & 11.68 & 51.9 & 99.1 & 73.0 & 28.1 & 68.9 & 15.2 & 67.5 & \phantom{0}9.3 & 31.4 \\
ViT-MLP          & 11.83 & 54.0 & 98.6 & 79.6 & 25.8 & 68.6 & \phantom{0}3.4 & 67.8 & 14.8 & 31.8 \\
ViT-Trans        & 11.61 & 76.2 & 66.6 & 43.8 & 28.9 & 66.7 & 28.7 & 59.6 & 18.9 & 29.7 \\
ViT-Combined     & 11.49 & 53.5 & 98.5 & 75.5 & 25.6 & 67.4 & \phantom{0}0.7 & 68.9 & \phantom{0}8.3 & 31.1 \\
\midrule
\multicolumn{11}{l}{\textit{Frozen foundation models}} \\
RadDINO-MLP      & \phantom{0}9.88 & 22.1 & 90.6 & 87.6 & 33.8 & 69.8 & \phantom{0}3.1 & 50.0 & 20.1 & 33.9 \\
RadDINO-Trans    & \phantom{0}9.63 & 64.2 & \textbf{100.0} & \textbf{94.1} & 29.0 & 68.6 & \phantom{0}9.0 & 50.8 & 16.4 & 33.8 \\
BiomedCLIP-MLP   & 10.47 & \phantom{0}0.7 & 97.6 & 82.8 & 36.4 & 68.2 & \phantom{0}2.4 & 69.8 & 12.8 & 30.7 \\
BiomedCLIP-Trans & \phantom{0}9.89 & 67.8 & 96.2 & 81.6 & 29.6 & 68.7 & 25.7 & 68.9 & 12.1 & 31.4 \\
SAM-Med2D-MLP    & 10.12 & \phantom{0}6.5 & 99.1 & 83.2 & 30.5 & 69.5 & \phantom{0}2.5 & 68.7 & 21.5 & 34.9 \\
SAM-Med2D-Trans  & 10.21 & 63.4 & 82.3 & 46.0 & 31.8 & 70.5 & 12.3 & 60.3 & 22.1 & 27.3 \\
\midrule
\multicolumn{11}{l}{\textit{Finetuned foundation models}} \\
RadDINO-MLP$^\dagger$   & 11.46 & 69.1 & 97.0 & 87.7 & 29.3 & 67.9 & 22.1 & 48.7 & 21.1 & 31.4 \\
RadDINO-Trans$^\dagger$ & 11.68 & \textbf{82.0} & 99.3 & 92.7 & 31.9 & 68.2 & 28.2 & 50.3 & 20.2 & 34.0 \\
\midrule
Chance & --- & --- & 50.0 & 20.0 & 33.3 & --- & --- & 50.0 & 20.0 & 33.3 \\
\bottomrule
\end{tabular}
\end{table}

Table~\ref{tab:main_results} reports both cohorts. In domain, CNN-MLP localizes best
(8.95\,px) and frozen RadDINO+Trans is competitive (9.63\,px), while finetuning degrades
both. ViT+Trans leads relational reasoning among from-scratch models (76.2\%) and
finetuned RadDINO+Trans overall (82.0\%); frozen models with MLP heads fail
(0.7--22.1\%) and recover to 63--68\% with a Transformer head. No configuration leads at all three levels.
Under zero-shot transfer, symmetry holds its band at 27.3--34.9\% while relative
position falls to 48.7--69.8\%, with frozen RadDINO+MLP at 50.03\,$\pm$\,0.04, and
orientation drops to 8.3--22.1\%. Localization does not transfer, with error rising to
66--71\,px, and permutation falls to 0.7--28.7\% with the ordering across configurations
changing rather than scaling down uniformly. Table~\ref{tab:sparvqa} reports the four
MLLMs: relative position sits at 49.5--50.1\% against 99\% for the encoders, orientation
at 8.5--20.2\%, and symmetry at 20.5--50.5\%, while organ size separates them from
7.7\% to 82.7\%.

\begin{table}[!t]
\centering
\caption{SPAR-VQA. Level~3 questions posed to four open-weight MLLMs (mean\,$\pm$\,std,
3 seeds, phrasing resampled per seed). All values in \% ($\uparrow$). Encoder rows give
the best and worst configuration from Table~\ref{tab:main_results} for reference.
Counting is omitted. Relative position is near ceiling for the encoders and at chance for
every MLLM.}
\label{tab:sparvqa}
\scriptsize
\setlength{\tabcolsep}{4pt}
\begin{tabular}{lcccccc}
\toprule
Model & Macro & RelPos & Orient & Symm & Size & Dist \\
& & \tiny(n=1162) & \tiny(n=1050) & \tiny(n=444) & \tiny(n=712) & \tiny(n=170) \\
\midrule
LLaVA-1.5-7B         & 34.9$\pm$0.4 & 50.1$\pm$1.1 & 20.2$\pm$0.2 & 50.2$\pm$0.4 & \phantom{0}7.7$\pm$0.4 & 46.5$\pm$3.1 \\
Qwen2-VL-2B-Instruct & 31.3$\pm$0.3 & 49.5$\pm$1.3 & 19.1$\pm$0.4 & 20.5$\pm$0.2 & 15.2$\pm$0.9 & 52.2$\pm$1.2 \\
MedGemma-4B-IT       & 46.8$\pm$2.4 & 50.0$\pm$0.0 & \phantom{0}8.5$\pm$2.0 & 50.5$\pm$0.0 & 71.3$\pm$10.3 & 53.7$\pm$0.3 \\
Qwen2-VL-7B-Instruct & 50.6$\pm$0.2 & 50.0$\pm$0.0 & 20.0$\pm$0.0 & 46.7$\pm$0.7 & 82.7$\pm$0.5 & 53.5$\pm$0.0 \\
\midrule
CNN-MLP (best enc.)  & 78.5 & 99.8 & 93.6 & 29.9 & 82.1 & 87.2 \\
ViT-Trans (worst enc.) & 55.1 & 66.6 & 43.8 & 28.9 & 53.0 & 83.2 \\
\midrule
Chance & --- & 50.0 & 20.0 & 33.3 & 33.3 & 50.0 \\
\bottomrule
\end{tabular}
\end{table}

\textit{The Architectural Sweep Establishes What Is Achievable.}
All 13 configurations see identical data, task heads, and training schedules. Permutation accuracy
ranges from 0.7\% for frozen BiomedCLIP+MLP to 82.0\% for finetuned RadDINO+Trans, and a
ViT trained from scratch on ImageNet initialization reaches 76.2\%. The task is learnable under our budget, so a low score reflects the representation and
its readout, not task difficulty.

\textit{Frozen Models Encode Position but Not Relations, and Pooling Hides It.} Frozen RadDINO localizes at 9.63\,px, close to the best configuration, but
reaches 22.1\% permutation accuracy with an MLP head; frozen BiomedCLIP reaches 0.7\%
and 67.8\% with a Transformer head on identical features. Positional information is
present while relational structure has to be computed outside the encoder. Neither
pretraining objective requires anatomical arrangement, since DINO optimizes view
invariance and CLIP optimizes text alignment, and BiomedCLIP's 15 million training images
confer no relational advantage over ImageNet initialization (ViT-MLP, 54.0\%), consistent
with UniBench~\cite{altahan2024unibench}. The two heads also differ in what they read. The MLP sees a pooled vector, the Transformer
sees the full set of tokens. So the jump from 0.7\% to 67.8\% shows what pooling throws
away, not what extra compute buys. The lesson for evaluation is the same. A pooled MLP
probe reports the relational structure as absent when it is still there in the tokens.
Symmetry stays at chance under both heads, so its failure is not about pooling.

\textit{Finetuning Trades Localization for Relations.}
Finetuning RadDINO improves permutation accuracy (+47 points with an MLP head, +18 with a
Transformer head) while degrading localization (9.88 to 11.46\,px for the MLP head) and
organ size. Performance shifts from favouring localization to favouring relations, which
is what we would expect if finetuning disrupted the metric features that coordinate
regression depends on, though we measure task accuracy rather than representation
geometry.
This matches the speciality--generality tradeoff reported for foundation model
finetuning~\cite{lin2023speciality}, where gains on a target task cost capabilities the
objective does not monitor. Task-specific finetuning is standard in medical imaging, so
this degradation would ordinarily go unmeasured.

\textit{High In-Domain Scores Can Reflect Canonical Anatomy.}
Relative position exceeds 98\% on WORD for most configurations, and CNN-MLP reaches
99.8\%. On TotalSegmentator the same weights score 50.0--69.8\%, with frozen RadDINO +MLP
at 50.03\,$\pm$\,0.04, indistinguishable from chance. Horizontal flipping prevents a
model from recalling the arrangement of a specific slice, but not from learning
regularities that hold across WORD. That the score survives one cohort and not another
suggests what we measured in domain was recall of canonical anatomy rather than a
measurement within the image. The collapse is selective. Relative position, orientation, and localization fall, while
distance and organ size hold. This is not a uniform drop from a broad feature shift.
Symmetry is the exception, but it was already at chance, so it had no room to fall.

\textit{Bilateral Symmetry Stays at Chance Throughout.}
All 13 configurations score 25.6--36.4\% on symmetry, a spread of 11 points against 50
for orientation, and neither architecture, pretraining scale, nor finetuning moves it. It stays in the same band on TotalSegmentator (27.3--34.9\%) while
every other probe shifts. The MLLMs stay near chance too. Two give nearly identical
answers across seeds, so they may be picking a fixed option rather than comparing the
kidneys. This is the one probe in SPAR with no fixed answer available, and nothing we evaluate
beats chance on it. We read this as a limit on what our readout heads recover, not on what
the representation holds. An untested head might succeed. The label may also be noisy: a
single axial slice cuts each kidney at a different level depending on where it falls and
how the patient lies, so the area ratio is not a clean target. A mask-input test would
show whether the label is recoverable at all, and we do not include one. Bilateral size
asymmetry is a screening criterion for renal artery
stenosis~\cite{artyszuk2022threshold}.

\textit{Encoder and System Competence Do Not Track Each Other.}
Relative position is near ceiling for the encoders on WORD and sits at 49.5--50.1\% for
all four MLLMs. Orientation reaches 93.6\% for CNN-MLP and 8.5--20.2\% for the MLLMs.
Organ size runs the other way, separating the MLLMs by 75 points across a range the
encoders cover within 40. The two levels therefore diverge in both directions. Our design
does not isolate where the difference arises, since several components sit between the two
measurements. A spatial capability reported at one level should not be assumed at the
other.
\section{Conclusion}

A spatial score depends on the representation, on the probe head reading it, and on the
cohort it was measured on. We changed each of these while holding the others fixed, and in
each case some capability that had looked established did not survive the change. A single
score therefore says less about a model than it appears to, and the cheapest way to find
out how much less is to move one of the three and measure again. On our probes, the ones that held up in domain
and then collapsed were those with an answer that is the same in most patients. Anatomy
makes that kind of probe easier to write, so a benchmark assembled from anatomical labels
may fill with them unintentionally. A probe with no canonical answer is worth including for
that reason alone, and it was also the one nothing we evaluated solved.
The encoders and the four MLLMs disagreed in both directions, and neither measurement
alone would have shown it. The same probes need testing in 3D, in other body regions, and in other modalities.


%
%

\clearpage
\section*{Supplementary Material}
\addcontentsline{toc}{section}{Supplementary Material}

This supplement reports per-organ and per-configuration results condensed in the main text. All values are computed on the same splits and seeds described in Section~3. Tables~\ref{tab:supp_l1} and~\ref{tab:supp_l2} give the per-organ breakdown behind the Level~1 and Level~2 aggregates in Table~\ref{tab:main_results}. Localization error is lowest for kidneys and highest for colon and intestine across every configuration. On permutation, frozen MLP heads fall to near zero on the esophagus while a Transformer head on the same features recovers it, which is the per-organ form of the pooling effect discussed in Section~4. Table~\ref{tab:supp_zs} gives the full per-configuration zero-shot transfer to TotalSegmentator, condensed to ranges in the main text; symmetry stays near chance for all 13 configurations while relative position and orientation vary across configurations rather than degrading uniformly.

\begin{table}[!]
\centering
\caption{Per-organ localization error (MED, px, mean over three seeds) on WORD. Lower is better. Kidneys are easiest; colon and intestine hardest across configurations.}
\label{tab:supp_l1}
\scriptsize
\setlength{\tabcolsep}{3pt}
\resizebox{\textwidth}{!}{%
\begin{tabular}{lcccccccccccc}
\toprule
Model & Liver & Spln & K-L & K-R & Stom & GB & Esph & Panc & Duod & Colon & Intes & Adrn \\
\midrule
\multicolumn{13}{l}{\textit{Trained from scratch}} \\
CNN-MLP & 8.3 & 8.0 & 7.2 & 7.1 & 10.3 & 9.1 & 8.6 & 9.1 & 8.1 & 11.5 & 10.9 & 9.5 \\
CNN-Trans & 10.7 & 10.6 & 8.3 & 8.8 & 14.1 & 11.4 & 9.0 & 11.9 & 9.6 & 16.8 & 15.0 & 12.1 \\
ViT-MLP & 11.6 & 10.6 & 8.5 & 8.4 & 14.3 & 11.3 & 7.3 & 11.1 & 10.0 & 17.6 & 15.3 & 11.2 \\
ViT-Trans & 11.2 & 10.2 & 8.0 & 8.3 & 13.8 & 10.7 & 10.1 & 10.8 & 9.6 & 17.6 & 16.1 & 11.0 \\
ViT-Comb & 11.1 & 10.0 & 8.2 & 8.2 & 13.5 & 11.6 & 8.5 & 10.9 & 9.6 & 17.2 & 15.1 & 11.0 \\
\midrule
\multicolumn{13}{l}{\textit{Frozen foundation models}} \\
RadDINO-MLP & 8.9 & 8.8 & 7.5 & 7.4 & 11.3 & 9.4 & 6.9 & 9.7 & 8.5 & 14.9 & 11.8 & 10.4 \\
RadDINO-Trans & 8.8 & 8.5 & 7.3 & 7.3 & 10.6 & 9.0 & 7.0 & 9.5 & 8.6 & 14.2 & 11.6 & 9.7 \\
BiomedCLIP-MLP & 9.8 & 9.6 & 8.1 & 7.7 & 11.8 & 10.4 & 8.3 & 10.0 & 9.0 & 15.1 & 12.5 & 11.1 \\
BiomedCLIP-Trans & 9.0 & 9.0 & 7.7 & 7.4 & 11.2 & 9.9 & 8.9 & 9.5 & 8.4 & 14.2 & 12.1 & 10.8 \\
SAM-Med2D-MLP & 8.9 & 9.2 & 7.7 & 8.0 & 11.8 & 9.5 & 8.0 & 9.7 & 8.8 & 14.7 & 12.3 & 10.9 \\
SAM-Med2D-Trans & 9.5 & 9.4 & 8.0 & 8.2 & 11.9 & 9.7 & 7.4 & 10.2 & 9.3 & 14.0 & 12.0 & 10.5 \\
\midrule
\multicolumn{13}{l}{\textit{Finetuned foundation models}} \\
RadDINO-MLP$^\dagger$ & 10.5 & 10.2 & 8.4 & 8.4 & 13.2 & 11.3 & 11.2 & 11.2 & 9.7 & 17.1 & 14.5 & 11.2 \\
RadDINO-Trans$^\dagger$ & 11.2 & 10.5 & 8.4 & 8.2 & 14.2 & 11.0 & 8.2 & 10.9 & 9.8 & 17.4 & 15.5 & 11.6 \\
\bottomrule
\end{tabular}}
\end{table}
\begin{table}[!]
\centering
\caption{Per-organ permutation accuracy (patch accuracy, \%, mean over three seeds) on WORD. Higher is better. Frozen MLP heads collapse on variable organs, most sharply the esophagus, while a Transformer head on the same features recovers them.}
\label{tab:supp_l2}
\scriptsize
\setlength{\tabcolsep}{3pt}
\resizebox{\textwidth}{!}{%
\begin{tabular}{lcccccccccccc}
\toprule
Model & Liver & Spln & K-L & K-R & Stom & GB & Esph & Panc & Duod & Colon & Intes & Adrn \\
\midrule
\multicolumn{13}{l}{\textit{Trained from scratch}} \\
CNN-MLP & 98.5 & 96.6 & 97.2 & 97.5 & 92.6 & 85.3 & 53.8 & 88.7 & 88.1 & 94.9 & 87.3 & 92.4 \\
CNN-Trans & 98.3 & 94.3 & 96.2 & 95.9 & 87.8 & 70.2 & 10.3 & 79.0 & 78.4 & 93.3 & 74.9 & 87.7 \\
ViT-MLP & 98.1 & 95.4 & 96.8 & 97.3 & 84.9 & 81.8 & 30.8 & 84.1 & 82.8 & 89.9 & 78.1 & 92.6 \\
ViT-Trans & 99.2 & 97.4 & 98.7 & 98.9 & 92.5 & 88.2 & 79.5 & 94.9 & 90.5 & 96.9 & 87.7 & 96.8 \\
ViT-Comb & 97.0 & 93.5 & 95.4 & 96.2 & 78.4 & 73.7 & 48.7 & 81.6 & 79.9 & 85.2 & 75.1 & 87.7 \\
\midrule
\multicolumn{13}{l}{\textit{Frozen foundation models}} \\
RadDINO-MLP & 92.6 & 83.7 & 73.9 & 70.9 & 75.1 & 55.3 & 12.8 & 62.9 & 61.6 & 82.3 & 60.0 & 85.4 \\
RadDINO-Trans & 98.0 & 94.1 & 93.7 & 94.4 & 91.3 & 79.8 & 82.1 & 88.2 & 83.2 & 95.1 & 84.0 & 93.2 \\
BiomedCLIP-MLP & 57.4 & 46.5 & 48.8 & 50.3 & 37.7 & 23.1 & 0.0 & 33.6 & 22.9 & 46.6 & 25.9 & 40.1 \\
BiomedCLIP-Trans & 98.4 & 95.0 & 95.6 & 95.1 & 90.0 & 79.4 & 74.4 & 92.1 & 84.7 & 96.4 & 85.9 & 93.5 \\
SAM-Med2D-MLP & 84.1 & 61.6 & 80.8 & 77.6 & 36.4 & 23.6 & 2.6 & 27.3 & 30.0 & 69.9 & 50.0 & 60.8 \\
SAM-Med2D-Trans & 98.4 & 93.9 & 96.4 & 97.2 & 87.6 & 82.4 & 53.8 & 87.0 & 82.6 & 93.5 & 82.9 & 93.5 \\
\midrule
\multicolumn{13}{l}{\textit{Finetuned foundation models}} \\
RadDINO-MLP$^\dagger$ & 98.9 & 97.2 & 97.9 & 98.5 & 92.5 & 87.6 & 59.0 & 90.6 & 89.5 & 95.8 & 88.7 & 94.4 \\
RadDINO-Trans$^\dagger$ & 99.6 & 97.7 & 98.5 & 99.0 & 96.2 & 92.5 & 82.0 & 96.8 & 93.9 & 98.5 & 92.5 & 96.2 \\
\bottomrule
\end{tabular}}
\end{table}
\begin{table}[t]
\centering
\caption{Full zero-shot transfer to TotalSegmentator, Level~3 probes (\%, mean over three seeds; RadDINO-Trans$^\dagger$ two seeds). No retraining. Symmetry stays near its 33.3\% chance level for every configuration while other probes shift.}
\label{tab:supp_zs}
\scriptsize
\setlength{\tabcolsep}{5pt}
\begin{tabular}{lccccc}
\toprule
Model & RelPos & Orient & Symm & Size & Dist \\
\midrule
\multicolumn{6}{l}{\textit{Trained from scratch}} \\
CNN-MLP & 68.8 & 9.1 & 30.9 & 38.4 & 48.2 \\
CNN-Trans & 67.5 & 9.3 & 31.4 & 34.3 & 49.1 \\
ViT-MLP & 67.8 & 14.8 & 31.8 & 32.7 & 46.9 \\
ViT-Trans & 59.6 & 18.9 & 29.7 & 34.7 & 54.2 \\
ViT-Comb & 68.9 & 8.3 & 31.1 & 31.5 & 49.1 \\
\midrule
\multicolumn{6}{l}{\textit{Frozen foundation models}} \\
RadDINO-MLP & 50.0 & 20.1 & 33.9 & 38.2 & 44.3 \\
RadDINO-Trans & 50.8 & 16.4 & 33.8 & 31.2 & 54.0 \\
BiomedCLIP-MLP & 69.8 & 12.8 & 30.7 & 37.2 & 52.7 \\
BiomedCLIP-Trans & 68.9 & 12.1 & 31.4 & 36.2 & 53.4 \\
SAM-Med2D-MLP & 68.7 & 21.5 & 34.9 & 32.3 & 43.7 \\
SAM-Med2D-Trans & 60.3 & 22.1 & 27.3 & 31.4 & 57.6 \\
\midrule
\multicolumn{6}{l}{\textit{Finetuned foundation models}} \\
RadDINO-MLP$^\dagger$ & 48.7 & 21.1 & 31.4 & 33.0 & 48.9 \\
RadDINO-Trans$^\dagger$ & 50.3 & 20.2 & 34.0 & 36.0 & 50.9 \\
\midrule
Chance & 50.0 & 20.0 & 33.3 & 33.3 & 50.0 \\
\bottomrule
\end{tabular}
\end{table}

\end{document}